\documentclass[reprint,showpacs,preprintnumbers,aps,amsmath,amssymb,prc,floatfix]{revtex4-1}
\usepackage{epsfig}
\usepackage{graphicx,color}
\usepackage{amsmath,amssymb,bm}

\begin{document}


\title{BCS Pairing Gap in the Infrared Limit of the Similarity Renormalization Group}

\author{S. Szpigel}
 \affiliation{Centro de R\'adio-Astronomia e Astrof\'isica Mackenzie, Escola de Engenharia,
Universidade Presbiteriana Mackenzie, S\~ao Paulo, Brazil}

\author{V. S. Tim\'oteo}
\affiliation{Grupo de \'Optica e Modelagem Num\'erica (GOMNI), Faculdade de Tecnologia, Universidade Estadual de Campinas - UNICAMP, Campinas, Brazil}

\author{E. Ruiz Arriola}
\affiliation{Departamento
  de F\'isica At\'omica, Molecular y Nuclear and Instituto Carlos I de
  Fisica Te\'orica y Computacional, \\  Universidad de Granada, E-18071
  Granada, Spain}%

\date{\today}

\begin{abstract}
Effective interactions have been used to compute the pairing gap for nuclear and neutron matter in several schemes. In this work we analyze the impact of phase-shift equivalent interactions within the BCS theory on the $^1S_0$-channel pairing gap for a translational invariant many-fermion system such as nuclear and neutron matter. We solve the BCS pairing gap equation on a finite momentum grid for a toy model separable gaussian potential in the $^1S_0$-channel explicitly evolved through the Similarity Renormalization Group (SRG) transformation and show that in the on-shell and continuum limits the pairing gap vanishes. For finite size systems the momentum is quantized and the on-shell limit is realized for SRG cutoffs comparable to the momentum resolution. In this case the pairing gap can be computed directly from the scattering phase-shifts by an energy-shift formula. While the momentum grid is usually used as an auxiliary way of solving the BCS pairing gap equation, we show that it actually encodes some relevant physical information, suggesting that in fact finite grids may represent the finite size of the system.
\end{abstract}

\maketitle


\section{\label{sec:introduction}Introduction}

The microscopic origin of pairing in nuclei was first driven by the
analogy to the BCS theory of
superconductivity~\cite{Bohr:1958zz}. Since then, the nature of
pairing correlations has provided a lot of insight in Nuclear
Physics~(for a review see e.g. Ref.~\cite{Dean:2002zx} and references
therein). A renaissance of the subject was experienced by the
production of heavy $N \sim Z$ nuclei which are achieved by Radioactive
Ion Beams~\cite{Satula:2005fy}.

Effective interactions have been used to compute the pairing gap for
nuclear and neutron matter in several schemes~\cite{Dean:2002zx}. There are claims in the literature that what determines the
pairing gap are the phase-shifts~\cite{Elgaroy:1997ti} and most
often the BCS approach is based on having the scattering phase-shifts as the basic
input of the calculation. On the other hand, there is an
arbitrariness in this procedure, as there are infinitely many
interactions leading to identical phase-shifts.

For instance, in the case of the pairing gap in the $^1S_0$-channel,
most $V_{\rm low-k}$ calculations provide a BCS gap
which has maximum at Fermi momentum $p_F \sim 0.8 ~ {\rm fm}^ {-1}$ and strength about
$3 ~ {\rm MeV}$~\cite{Sedrakian:2003cc,Hebeler:2006kz}. In medium $T$-matrix has been used to provide an improvement on the standard BCS
theory~\cite{Bozek:2002ry} yielding a 30\% reduction in the $^1S_0$
pairing gap. {\it Ab initio}
calculations, however, may provide completely different
results~\cite{Gezerlis:2009iw,Gandolfi:2009tv,Baldo:2010du}. Thus, it is disconcerting that the BCS gap is so different and so much
scheme dependent. In this paper we analyze these ambiguities.

The BCS state provides a paring gap given by
\begin{eqnarray}
\Delta ({\bf k}) = -\frac12 \int \frac{d^3 {\bf p}}{(2\pi)^3} \frac{V({\bf k},{\bf p}) \Delta({\bf p})}{E({\bf p})} \; ,
\end{eqnarray}
with

\begin{equation}
E({\bf p})^2 = [({\bf p}^2-p_F^2)/(2M))^2 + \Delta({\bf p})]^2 \;,
\end{equation}
\noindent
where $M$ is the nucleon mass. After the partial-wave decomposition in terms of the tensor spherical harmonics ${\cal Y}_{LS}^{JM}$ due to the $J=L+S$ coupling~\cite{Dean:2002zx},
\begin{eqnarray}
V^S ({\bf p'},{\bf p}) &=& \frac{4\pi^2}{M}
\sum_{JMLL'} {\cal Y}_{LS}^{JM} (\hat p') V_{LL'}^{JS} (p',p)
{{\cal Y}_{L'S}^{JM}}^\dagger (\hat p)\; , \nonumber\\
\Delta^S ({\bf p}) &=& \sum_{JML} {\cal Y}_{LS}^{JM} (\hat p) \Delta_{L}^{JS} (p) \; ,
\end{eqnarray}
we have,
\begin{eqnarray}
\Delta_L^J (p) = -\frac1{\pi} \int_0^\infty dk ~k^2
\sum_{L'} \frac{V_{L,L'}^J (k,p) \Delta_{L'}^J (p)}{M E(p)} \; ,
\label{pwgap}
\end{eqnarray}
which is the generalized gap equation in all channels. These equations are solved
iteratively until convergence is achieved (for several strategies see
e.g.~\cite{khodel1996solution}). The pairing gap in a given
channel is defined as $\Delta_F=\Delta(p)|_{p=p_F}$.

We aim to analyze numerically the behavior of the pairing
gap as a function of the SRG cutoff towards the infrared limit. These
are demanding calculations, particularly with interactions having a
strong short distance repulsive core which provide long momentum
tails.

\section{BCS pairing gap with a toy model}

For our illustration purposes, in the present study we consider the toy model separable gaussian potential in the $^1S_0$-channel discussed in
our previous works~\cite{Arriola:2013era,Arriola:2014aia,Arriola:2014fqa,Arriola:2013gya},
\begin{equation}
V(p, p') = C~ g_L(p) g_L(p') \; ,
\label{gaupot}
\end{equation}
\noindent
with $g_L(p) = e^{-p^2/L^2}$. The toy model separable gaussian potential considerably reduces the computational effort for the SRG numerical treatment in the infrared limit, since the long momentum tails are suppressed from the start. The parameters $C$ and $L$ are determined from the solution of the Lippmann-Schwinger (LS) equation for the on-shell $T$-matrix by fitting the experimental values of the parameters of the Effective Range Expansion (ERE) to second order, i.e. the scattering length $a_0$ and the effective range $r_e$. Namely, we solve the LS equation for the $T$-matrix with the toy model potential,
\begin{eqnarray}
T(p,p';E)&=&V(p,p')\nonumber\\
&+&\frac{2}{\pi} \int_{0}^{\infty} \; dq \; q^2
\frac{V(p,q)}{E-q^2+i \; \epsilon} T(q,p';E),
\end{eqnarray}
\noindent
where $E$ is the scattering energy, and match the resulting on-shell $T$-matrix to the ERE expansion,
\begin{eqnarray}
T^{-1}(k,k;k^2)&=&-\left[-\frac{1}{a_0}
+\frac{1}{2}~r_e~k^2 + {\cal O}(k^4)-i~k \right]\nonumber\\
&=&-\left[k~{\rm cot}~\delta(k)-i~k \right]\; ,
\label{eq:ERE}
\end{eqnarray}
\noindent
where $k=\sqrt{E}$ is the on-shell momentum in the center of mass (CM) frame and $\delta(k)$ stands for the phase-shifts. In the case of the toy model separable gaussian potential it is straightforward to determine the phase-shifts from the solution of the LS equation for the
$T$-matrix using the {\it ansatz} $T(p,p';k^2) = g_L(p) ~t(k)~ g_L(p')$, where $t(k)$ is called the reduced on-shell $T$-matrix. This leads to
a simple relation (valid for separable potentials only),
\begin{eqnarray}
 k \cot \delta(k)= - \frac{1}{V(k,k)}\left[1-\frac{2}{\pi}
{\cal P}\int_0^\infty dq~q^2 \frac{V(q,q)}{k^2-q^2} \right], 
\label{phasetoy}
\end{eqnarray}
\noindent
where ${\cal P}$ denotes the Cauchy principal value. The phase-shifts calculated with the toy model separable gaussian potential are rather reasonable for CM on-shell momenta $k$ below $ \sim 1 ~{\rm fm}^{-1}$, when compared to the results obtained for realistic $NN$ potentials.

The BCS gap equation in the $^1S_0$-channel,

\begin{eqnarray}
\Delta (k) = -\frac1{\pi} \int_0^\infty  dp ~p^2
\frac{V (k,p) \Delta(p)}{M E(p)} \; ,
\label{eq:BCS}
\end{eqnarray}
\noindent
is also readily solved by taking the {\it ansatz} $\Delta(k) = \Delta_0 g(k)$, where $\Delta_0$ satisfies the implicit equation
\begin{eqnarray}
1 = -\frac1{\pi} \int_0^\infty dp~p^2 \frac{C [g(p)]^2}{M E(p)}\; .
\end{eqnarray}
\noindent
where $E(p)=[((p^2-p_F^2)/(2M))^2 + \Delta_0^2 g(p)^2]^{1/2}$.

\begin{figure}[t]
\begin{center}
\includegraphics[width=0.85\linewidth]{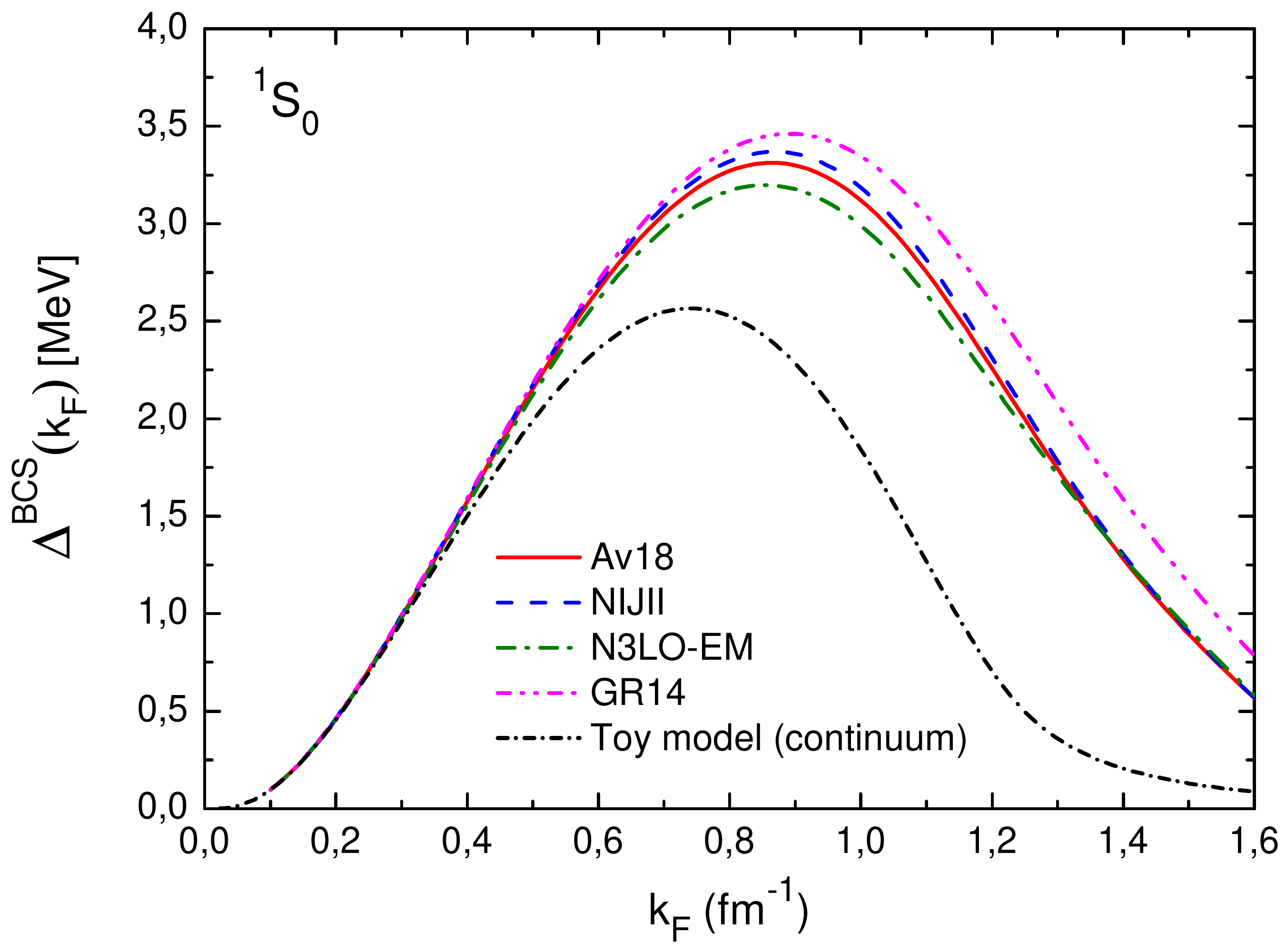}
\end{center}
\caption{BCS pairing gap for the toy model potential in the $^1S_0$-channel 
  compared to the results obtained for the realistic potentials AV18~\cite{Wiringa:1994wb},
  NijII~\cite{Stoks:1994wp}, N3LO-EM~\cite{Entem:2003ft} and
  GR14~\cite{Perez:2014yla}.}
\label{fig:BCS-toy-vs-pots}
\end{figure}

In Fig.~\ref{fig:BCS-toy-vs-pots} we show the BCS gap for the toy model potential, $\Delta_F \equiv \Delta_0 g(p_F)$, compared to the results obtained for the realistic $NN$ potentials AV18~\cite{Wiringa:1994wb}, NijII~\cite{Stoks:1994wp},
N3LO-EM~\cite{Entem:2003ft} and GR14~\cite{Perez:2014yla}. As one can see, the toy model potential yields a sufficiently reasonable behavior.

Most analysis based on the BCS approach usually end here. However, it is important to note that the only physical input information in these calculations
is contained in the phase-shifts but {\it not} in the potentials. As it
is well known, this generates some off-shell ambiguity which cannot be
directly related to measurable physical
information. This off-shellness corresponds
to the non-diagonal matrix elements of the potential. In
fact, one can make a unitary transformation $V \to U V U^\dagger$ such
that phase-shifts remain invariant (see e.g. Ref.~\cite{Delfino:2006}). In the following sections we will show
that the BCS pairing gap can in principle depend strongly on this
unitary transformation and hence on the off-shellness.

\section{BCS pairing gap equations on a finite momentum grid}

The BCS equations can be solved numerically on an $N$-dimensional
momentum grid, $p_1 < \dots p_N $ ~\cite{Szpigel:2010bj} by
implementing a high-momentum ultraviolet (UV) cutoff, $p_{\rm
  max}=\Lambda$, and an infrared (IR) momentum cutoff $p_{\rm min} =
\Delta p$. The integration rule becomes
\begin{eqnarray}
\int_{\Delta p}^\Lambda dp f(p) \to \sum_{n=1}^N w_n f(p_n) \, .
\end{eqnarray}

The completeness relation in discretized momentum-space reads:
\begin{equation}
1=\frac{2}{\pi}\sum_{n=1}^N w_n p_n^2 | p_n \rangle \langle p_n | \, .
\end{equation}
\noindent
By inserting this into Eq.~(\ref{pwgap}) and defining the matrix-elements of the potential as $V(p_n,p_m) \equiv \langle p_n |V | p_m \rangle$, we obtain the BCS equation on the finite momentum grid
\begin{eqnarray}
\Delta (p_n) = -\frac{2}{\pi} \sum_{k=1}^N w_k p_k^2 \frac{V(p_n,p_k)\Delta (p_k)}{2 ME (p_n)} \;,
\end{eqnarray}
where $ 2 M E (p_n) = \sqrt{(p_n^2-p_F^2)^2+ 4 M^2 \Delta
  (p_n)^2}$. Of course, on the grid the Fermi momentum must also
belong to the grid ($p_F= p_m$).

While the momentum grid is usually regarded as an auxiliary
element for solving the BCS gap equation, we will show that it actually encodes some
relevant physical information, suggesting that in fact finite grids may
represent the finite size of the system. Moreover, we will show that using the
inherent arbitrariness of the off-shellness in the potential one may
get a large variety of results. As a matter of fact, we will present a
scheme which is free of any off-shell ambiguities, and for this scheme
the continuum limit is shown to produce a vanishing BCS gap for an
infinitely large system.

\section{Phase equivalent interactions and the on-shell limit}

\def\tr{{\rm Tr}}

Quite generally, for a given hamiltonian we can always perform a
unitary transformation $H \to U H U^ \dagger $ keeping the phase-shifts
invariant. On a finite momentum grid the definition
of the phase-shift must be specified, since on the one hand one
replaces the scattering boundary conditions with standing waves
boundary conditions and on the other hand one wants to preserve the
invariance under unitary transformations on the grid.

The unitary transformation $U$ can be quite general, and for our study
we will generate them by means of the so-called similarity
renormalization group (SRG), proposed by Glazek and
Wilson~\cite{Glazek:1993rc,Glazek:1994qc} and independently by
Wegner~\cite{wegner1994flow} who showed how high- and low-momentum degrees of
freedom can decouple while keeping scattering equivalence.

The general SRG equation is given by~\cite{Kehrein:2006ti},
\begin{eqnarray}
\frac{d H_s}{ds} = [[ G_s, H_s],H_s] \; ,
\label{eq:SRG}
\end{eqnarray}
and supplemented with a generator $G_s$ and an initial condition at
$s=0$, $H_0$. This correspond to a one-parameter operator evolution
dynamics and, as it is customary, we will often switch to the SRG
cutoff $\lambda=s^{-1/4}$ which has momentum dimensions. The generator
$G_s$ can be chosen according to certain requirements, and
three popular choices are the kinetic energy
$T$~\cite{Glazek:1994qc} (Wilson-Glazek generator), the diagonal part
of the hamiltonian ${\rm Diag}(H)$ ~\cite{wegner1994flow} (Wegner
generator) or a block-diagonal (BD) generator $P H_s P + Q H_s Q $
where $P+Q=1 $ are orthogonal projectors $P^2 = P$, $Q^2 = Q$ ,
$QP=PQ=0$, for states below and above a given
momentum scale~\cite{Anderson:2008mu}.

On the finite momentum grid the SRG equations become a set of
non-linear coupled differential equations. For the Wegner generator, which will be
taken here for definiteness, the equations take a quite simple form
\begin{eqnarray}
\frac{d H_s (p_n,p_k)}{ds} =\frac2{\pi} \sum_{m}  H_s (p_n,p_m) w_m p_m^2 H_s (p_m,p_k) \nonumber \\ \times \left[H_s (p_n, p_n)
- 2 H_s (p_m, p_m) + H_s (p_k,p_k) \right] \; .
\label{SRGeq}
\end{eqnarray}

The BCS pairing gap equation for the SRG-evolved hamiltonian can be written as
\begin{equation}
\Delta_\lambda (p_n) = -\sum_{k=1}^N \frac{[H_\lambda (p_n,p_k) - p_n^2 \delta_{nk}] \Delta_\lambda (p_k)}{2 ME_\lambda (p_n)} \; ,
\end{equation}
Clearly, the BCS pairing gap becomes a function of the SRG cutoff $\lambda$.

The limit $\lambda \to 0$ corresponds to an infrared fixed-point of the SRG evolution, at which the Hamiltonian becomes a
diagonal matrix ~\cite{Arriola:2013gya},
\begin{equation}
\lim_{\lambda \to 0} H_\lambda (p_n,p_m)=P_n^2 \;,
\end{equation}
where $P_n^2$ the $n-$th ordered eigenvalue of the Hamiltonian. In this limit the potential also becomes diagonal, and hence all off-shellness is eliminated. Thus, in the SRG infrared limit $\lambda \to 0$ we get an on-shell interaction. An important result derived in Ref.~\cite{Arriola:2014fqa} is the energy-shift formula,
\begin{eqnarray}
\delta^{\rm ES} (p_n) = - \pi \lim_{\lambda \to 0} \frac{H_{n,n}^{G,\lambda}-p_n^2}{2 w_n p_n}=- \pi  \frac{P_n^2-p_n^2}{2 w_n p_n} \; ,
\label{eq:es-ps}
\end{eqnarray}
which provides phase-shifts that remain constant along the SRG-trajectory, i.e.
\begin{equation}
\delta_\lambda^{\rm ES} (p_n) = \delta_\infty^{\rm ES} (p_n) = \delta_0^{\rm ES} (p_n) \; .
\end{equation}
Furthermore, in the SRG infrared limit $\lambda \to 0$ the BCS pairing gap equation on the grid is determined by the energy-shift at the
Fermi surface for on-shell interactions,
\begin{eqnarray}
\Delta^{ES}_\lambda (p_n)&=&\lim_{\lambda \to  0} \Delta_\lambda (p_n) = w_n \frac{p_n \delta^{\rm ES} (p_n)}{\pi M}  \; .
\end{eqnarray}
It is important to note that the integration
weights $w_n$ appear explicitly in the formula, and in the continuum
limit $N \to \infty$ they vanish as $w_n= {\cal O}(1/N)$. Therefore, if we denote by $\Delta p_F \equiv
w_F $ the integration weight corresponding to the Fermi momentum, in
the continuum limit the BCS pairing gap becomes
\begin{eqnarray}
\Delta_F \sim \Delta p_F \frac{p_F \delta (p_F)}{\pi M} \; .
\label{eq:DeltaPF}
\end{eqnarray}
whenever $\Delta_F > 0$ and zero otherwise. This is our main result. One should also note that while the shape is rather universal, the strength is related
to $\Delta p_F$ which ultimately depends on the system size $R$ and
geometry, and for large systems $\Delta p_F= {\cal O}(1/R)$.

\begin{figure}[h]
\begin{center}
\includegraphics[width=0.47\linewidth]{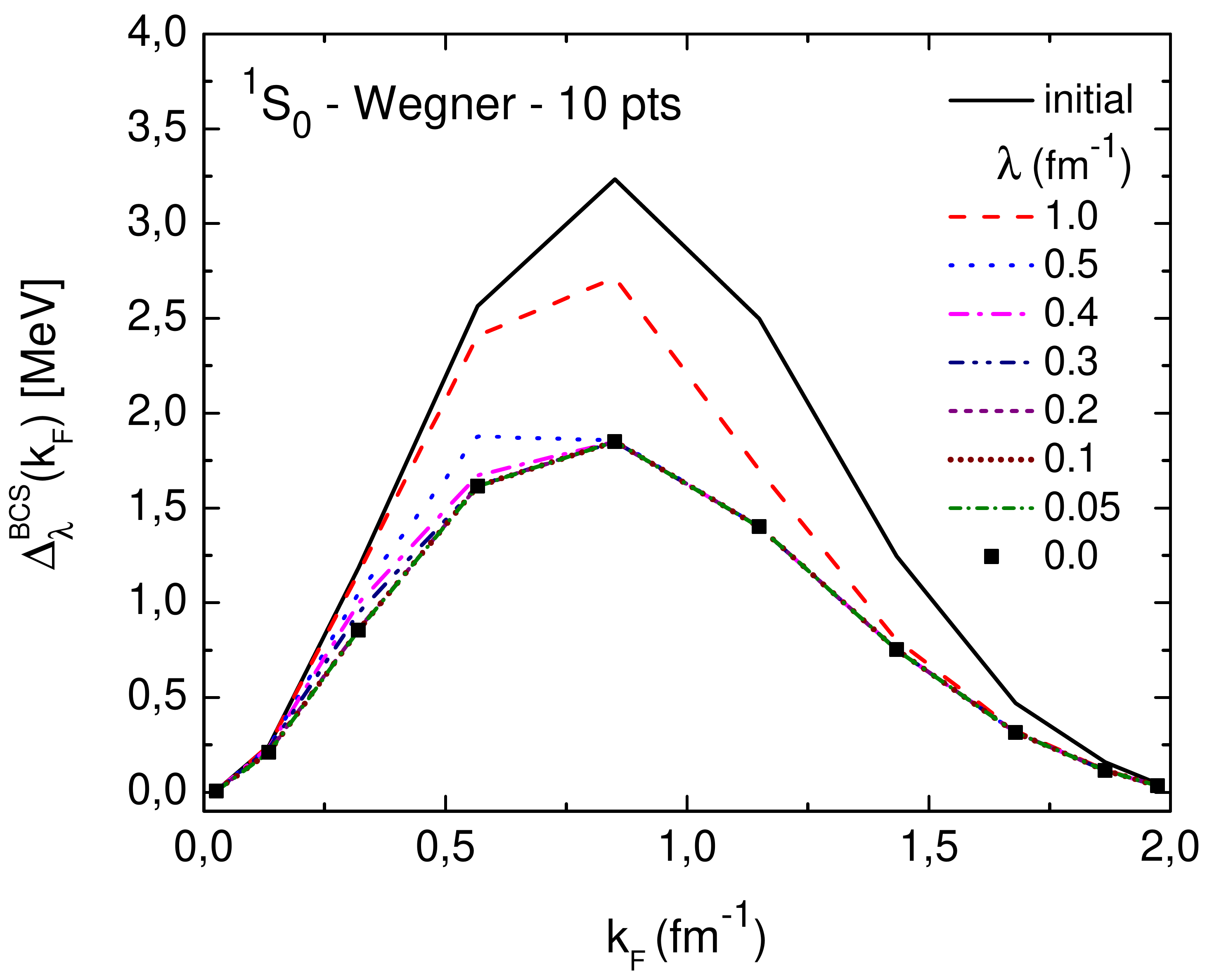}
\includegraphics[width=0.51\linewidth]{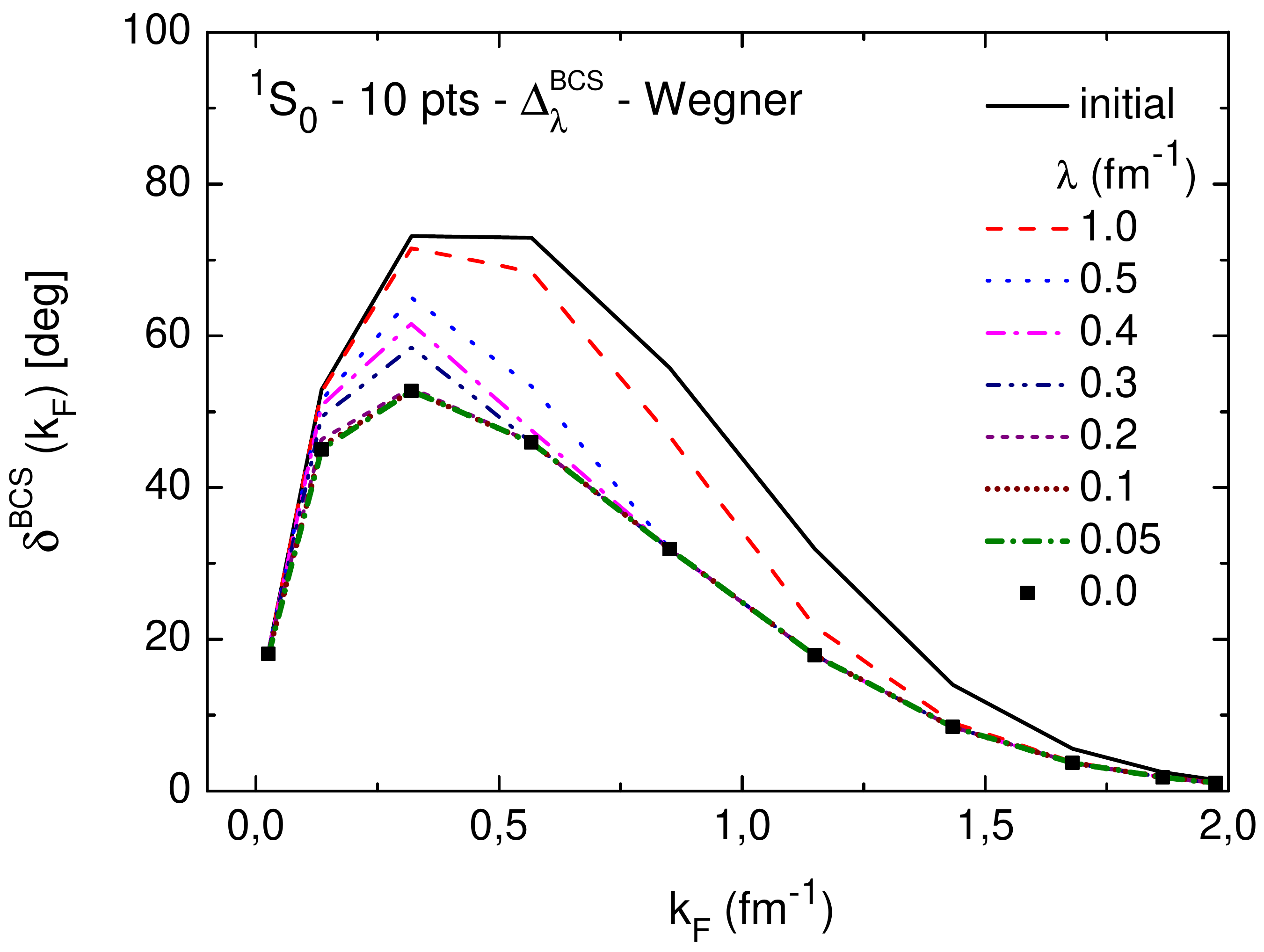} \\
\includegraphics[width=0.47\linewidth]{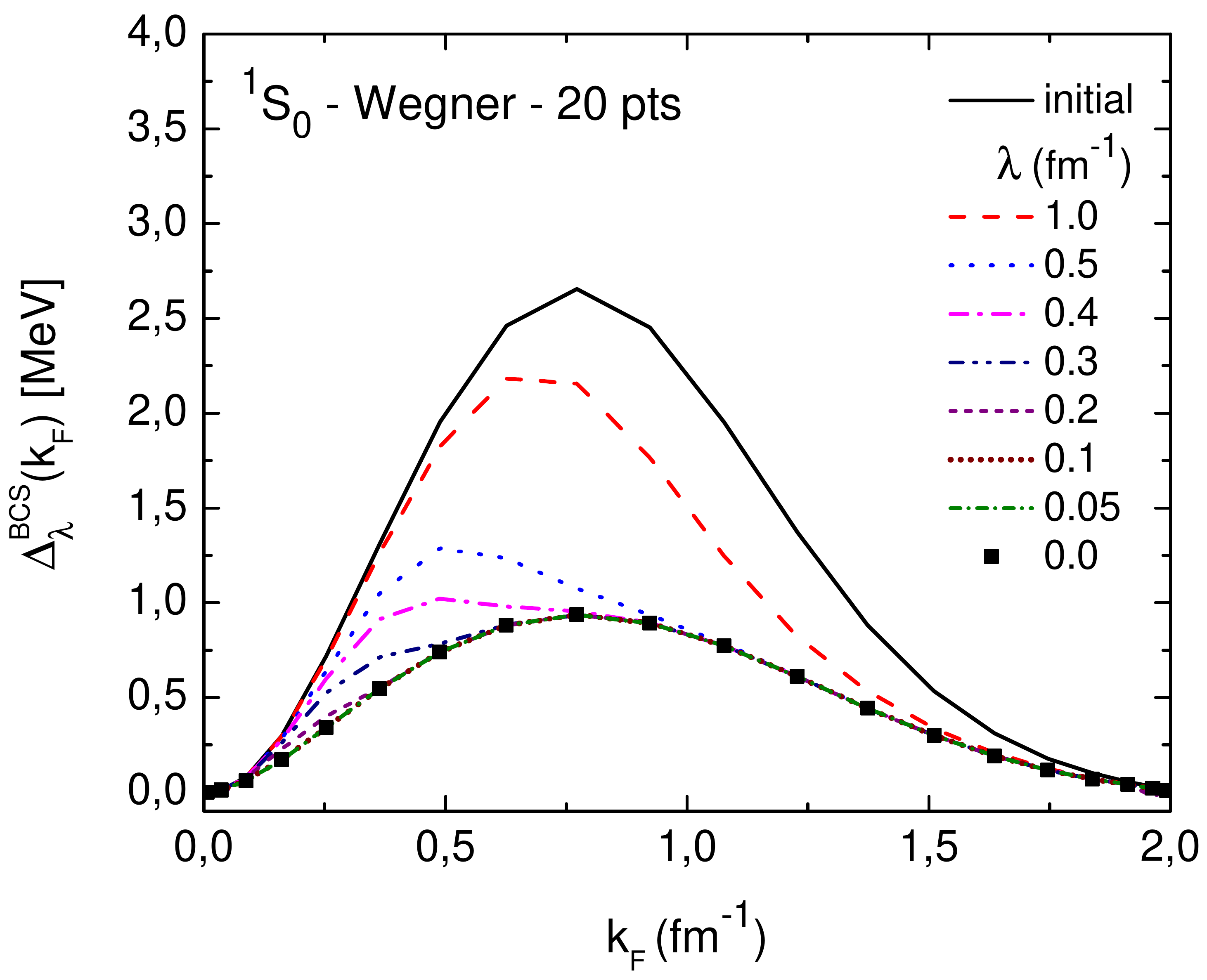}
\includegraphics[width=0.51\linewidth]{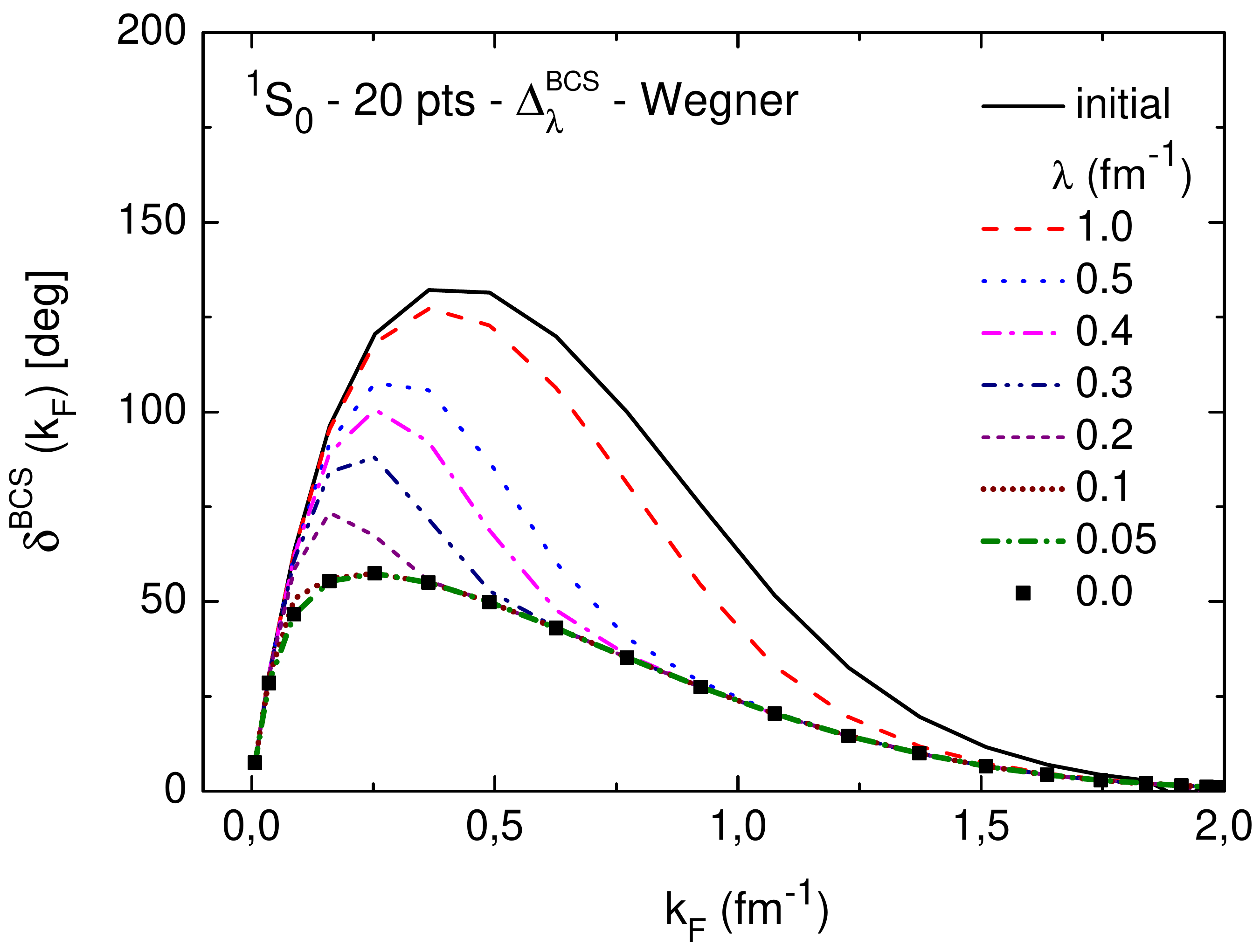} \\
\includegraphics[width=0.47\linewidth]{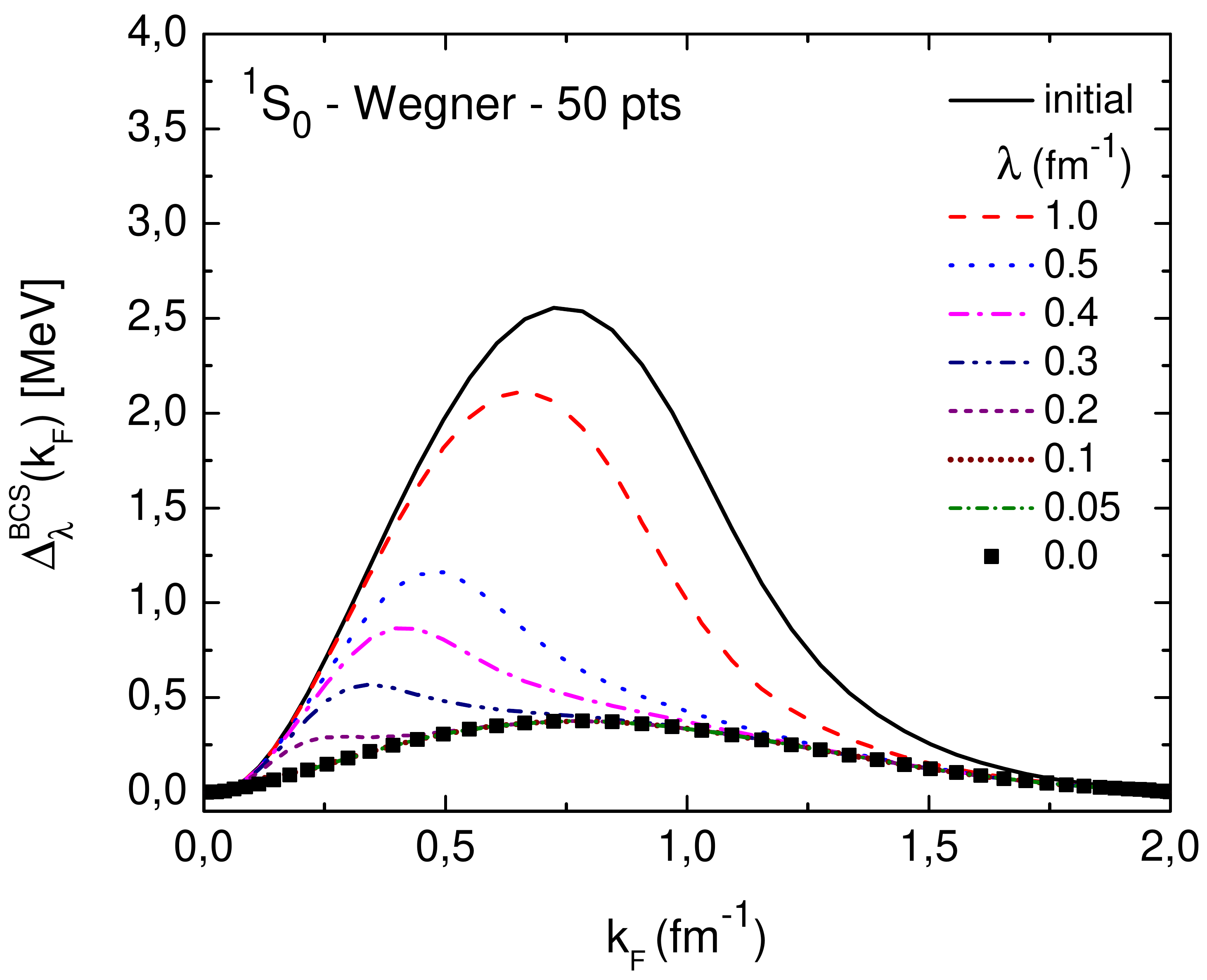}
\includegraphics[width=0.51\linewidth]{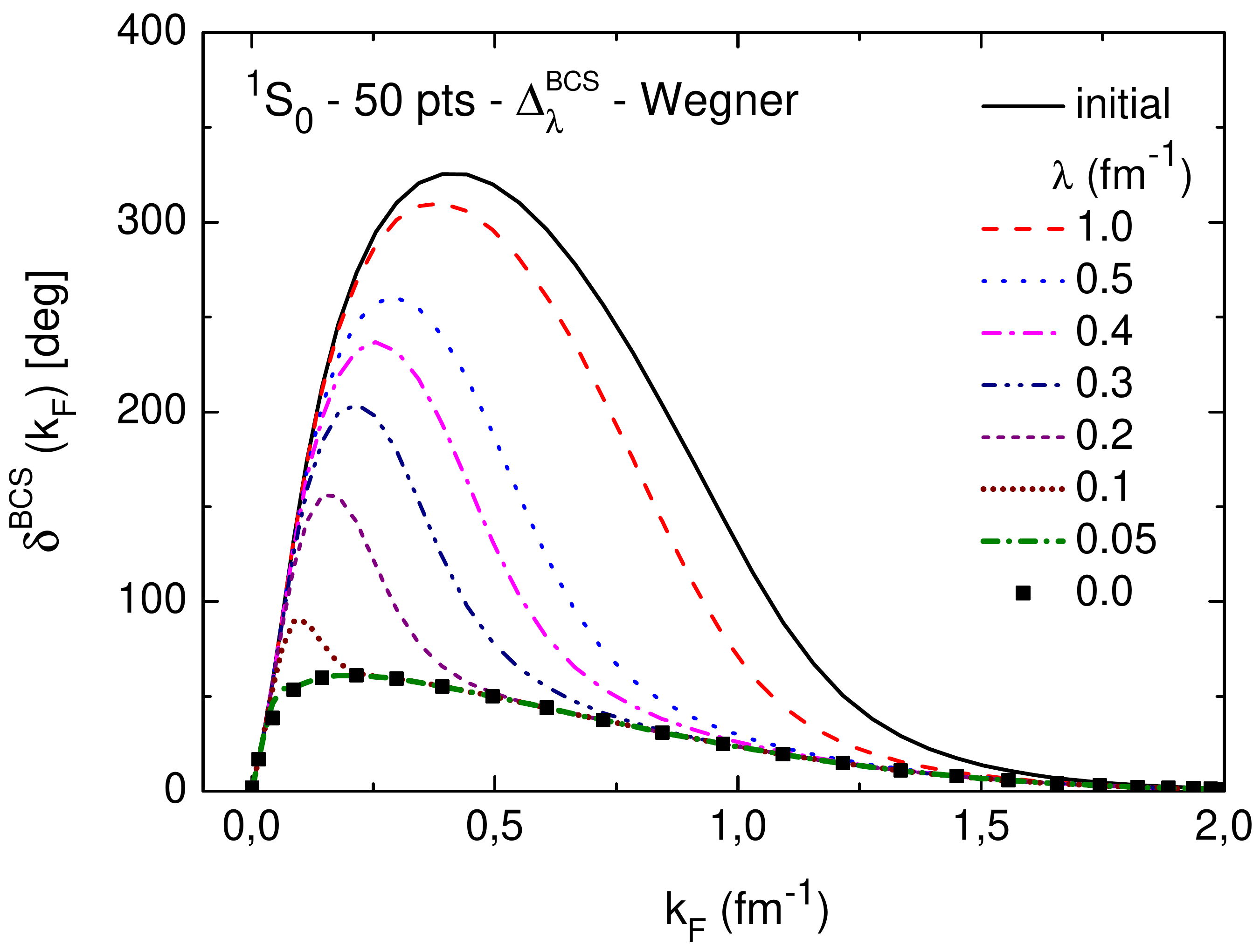} \\
\includegraphics[width=0.47\linewidth]{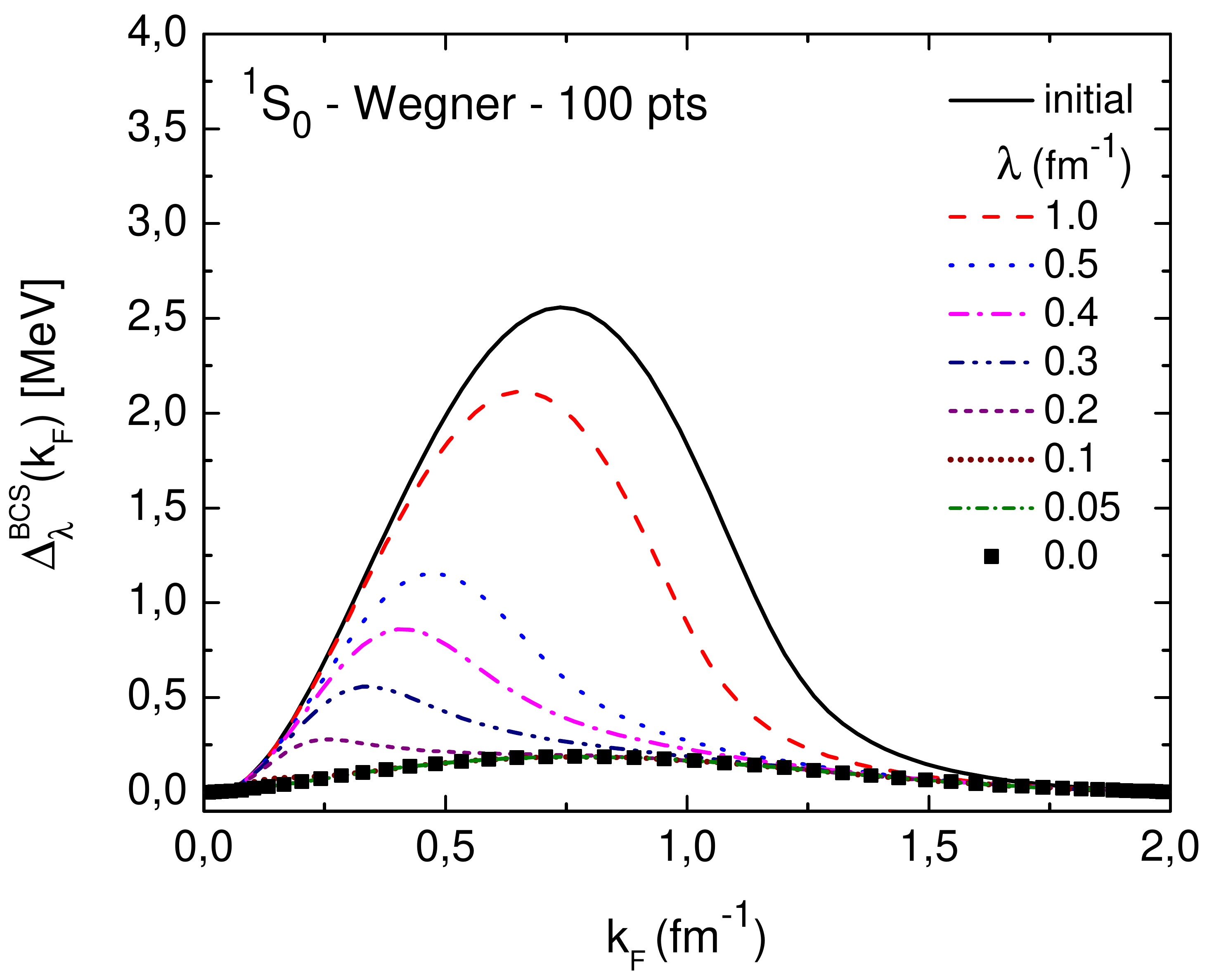}
\includegraphics[width=0.51\linewidth]{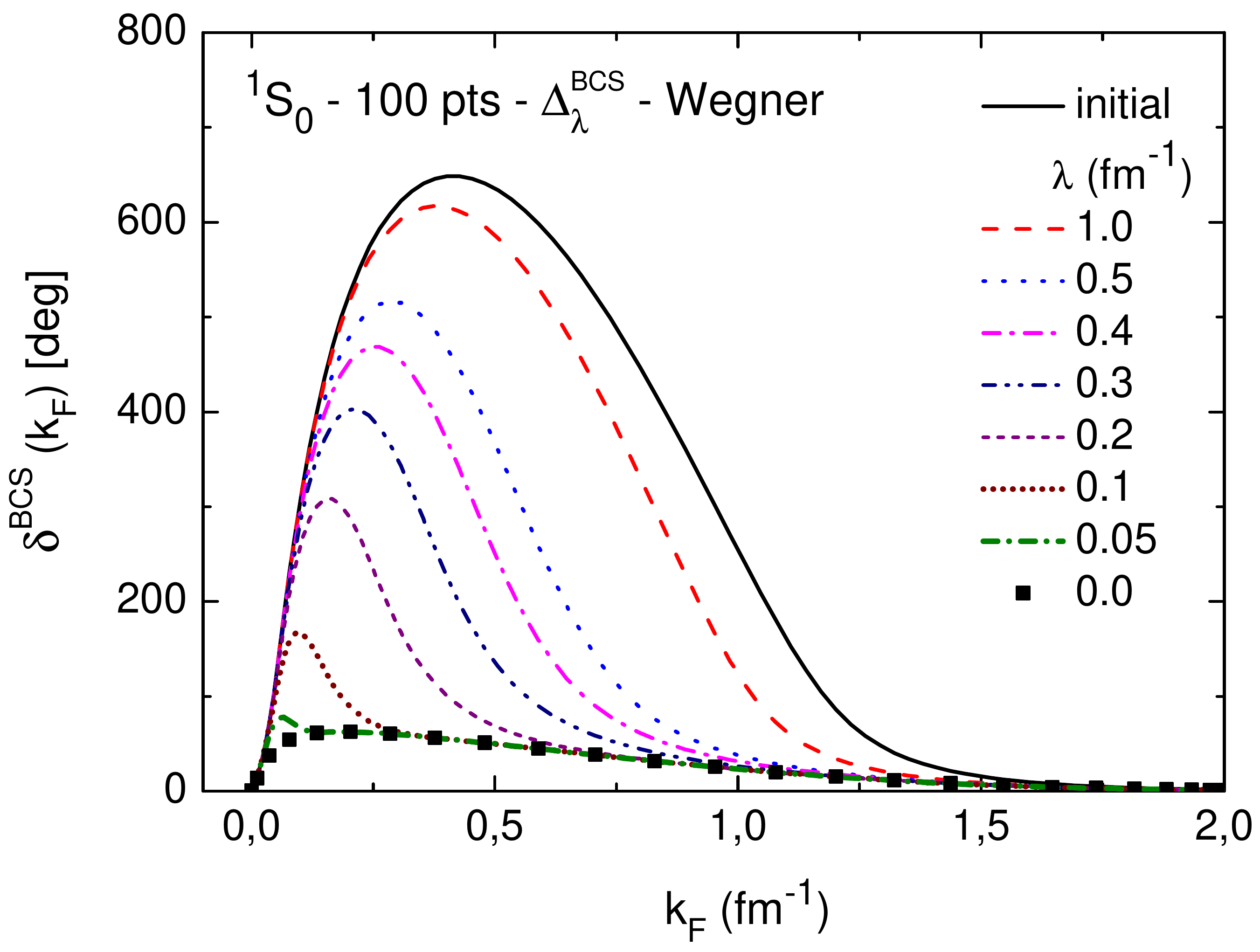}
\end{center}
\caption{SRG evolution of the BCS pairing gap for the toy model potential in the $^1S_0$-channel (left panels), obtained by solving the BCS equation on the finite momentum grid, and the corresponding BCS phase-shifts (right panels).}
\label{fig:BCS-SRG}
\end{figure}

In Fig.~\ref{fig:BCS-SRG} we show the SRG evolution of the BCS pairing gap $\Delta_\lambda^{\rm BCS} (k_F)$ for the toy model potential in the $^1S_0$-channel,  obtained by solving the BCS equation on the finite momentum grid for different number of grid points.  Alternatively, we
illustrate the scaling behavior of the BCS pairing gap by defining the ``BCS phase-shifts'' as
\begin{eqnarray}
\delta_\lambda^{\rm BCS} (k_F) = \frac{ \Delta_\lambda^{\rm BCS} (k_F) \pi M}{w_F k_F} \; ,
\label{eq:BCS-ps}
\end{eqnarray}
\noindent
which, as expected, converge to the phase-shifts obtained from the
energy-shift formula in the limit $\lambda \to 0$.  We remind that, as pointed out and illustrated in
Ref.~\cite{Arriola:2014aia}, the phase-shifts calculated through the solution of the LS equation {\it does not} fulfill the phase-invariance on the finite momentum grid, but only in the continuum limit, i.e. for $N \to \infty$. On the other hand, the phase-shifts remain invariant if we consider the energy-shift
definition~\cite{Arriola:2014aia}.

\begin{figure}[h]
\begin{center}
\includegraphics[width=0.66\linewidth]{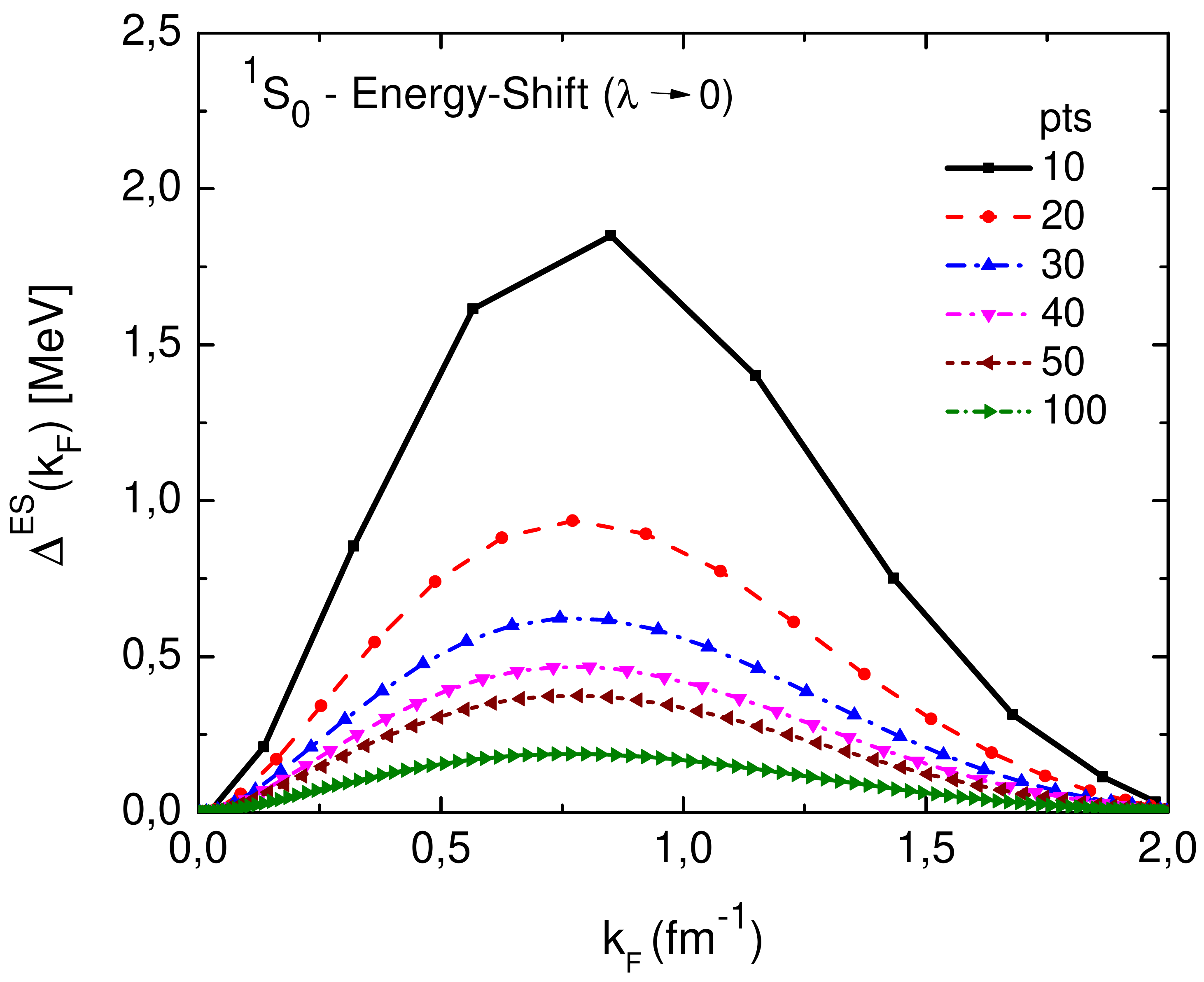}\\
\vspace{0.1cm}
\includegraphics[width=0.66\linewidth]{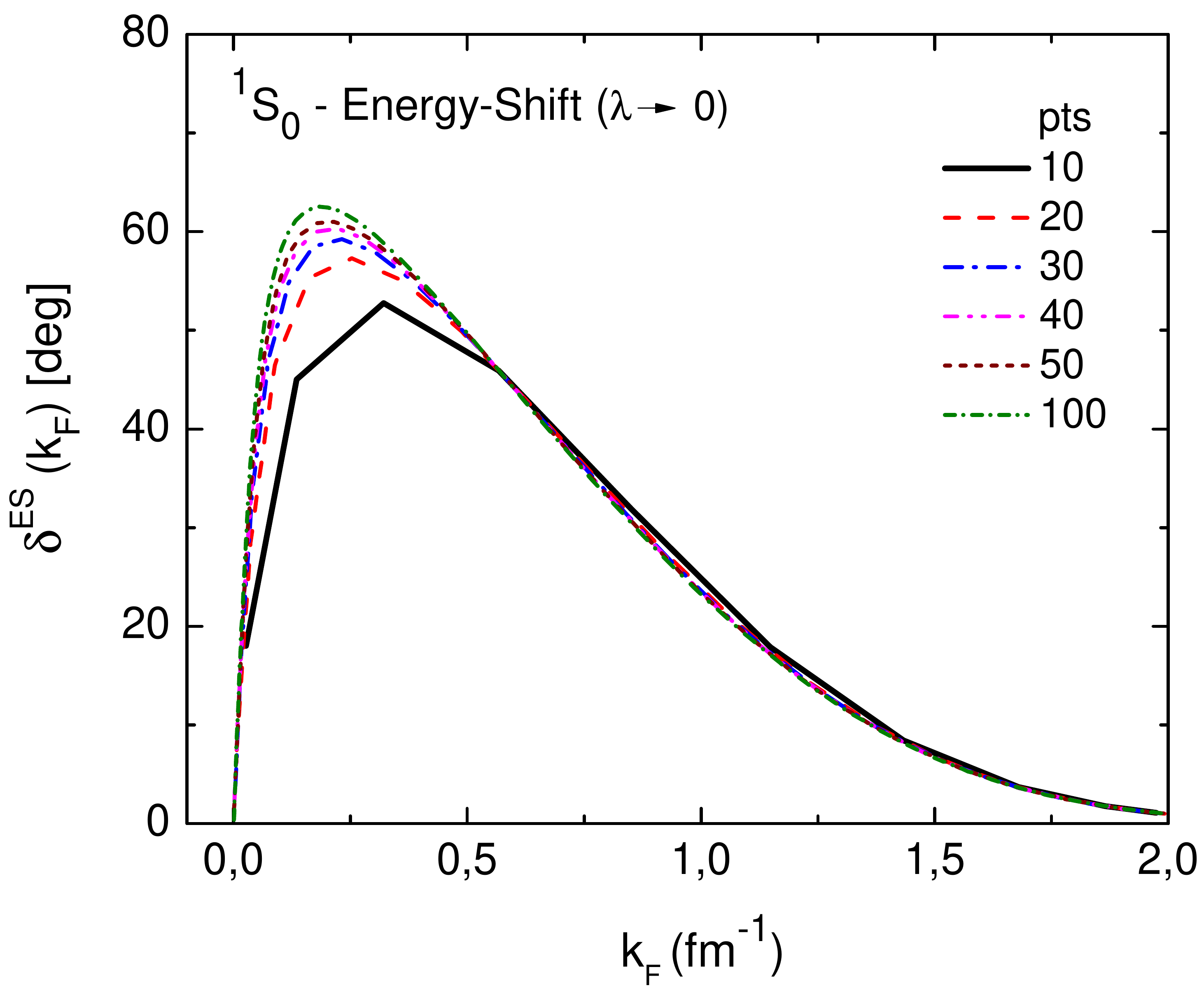}
\end{center}
\caption{BCS pairing gap for the toy model potential in the $^1S_0$-channel (top panel) and the corresponding phase-shifts (bottom panel), obtained from the energy-shift formula.}
\label{fig:pairing-simcut-0}
\end{figure}

In Fig.~\ref{fig:pairing-simcut-0} we show the BCS pairing gap $\Delta^{\rm ES} (k_F)$ and the corresponding phase-shifts $\delta^{\rm ES} (k_F)$,
in the infrared limit $\lambda \to 0$, obtained from the energy-shift formula. As one can see,
when we take the SRG infrared limit the BCS pairing gap vanishes in the strict continuum limit.

\section{Final Remarks}

In this work we have explored the freedom on reducing the off-shellness of the $NN$ interaction in the $^1S_0$ channel through the SRG evolution towards the infrared limit as a way to analyze the BCS pairing gap. Remarkably, we find that there is an on-shell regime where the BCS pairing gap can be directly determined by the $NN$ phase-shifts. We have verified by explicit numerical calculations that in the infrared ($\lambda \to 0$) and continuum ($N \to \infty$) limits the pairing gap vanishes, suggesting that finite momentum grids may represent the finite size of the system.

\begin{acknowledgments}
S.S. was supported by FAPESP, V.S.T. by FAEPEX/PRP/UNICAMP, FAPESP and CNPq and E.R.A. by Spanish Mineco (grant FIS2014-59386-P) and Junta de Andalucia (grant FQM225). Computational power provided by FAPESP (grant 2011/18211-2).
\end{acknowledgments}

\end{document}